\documentclass[12pt]{article}
\usepackage{psfig,epsfig}
\def\bd{\begin{document}} \def\ed{\end{document}} \def\r#1{$^{[#1]}$}
\def\bmp{\begin{minipage}} \def\emp{\end{minipage}}
\def\bcc{\begin{center}} \def\ecc{\end{center}}     \def\npg{\newpage}
\def\beq{\begin{equation}} \def\eeq{\end{equation}} \def\hph{\hphantom}
\def\beqr{\begin{\eqnarray}} \def\eeqr{\end{\eqnarray}}
\def\n{\noindent} \def\ni{\noindent} \def\pa{\parindent}
\def\hs{\hskip} \def\vs{\vskip} \def\hf{\hfill} \def\ej{\vfill\eject}
\def\cl{\centerline} \def\ob{\obeylines}  \def\ls{\leftskip}
\def\underbar#1{$\setbox0=\hbox{#1} \dp0=1.5pt \mathsurround=0pt
   \underline{\box0}$}   \def\ub{\underbar}    \def\ul{\underline}
\def\f{\left} \def\g{\right} \def\e{{\rm e}} \def\o{\over}
\def\vf{\varphi} \def\pl{\partial} \def\cov{{\rm cov}} \def\ch{{\rm ch}}
\def\la{\langle} \def\ra{\rangle} \def\EE{e$^+$e$^-$}
\def\bitz{\begin{itemize}} \def\eitz{\end{itemize}}
\def\btbl{\begin{tabular}} \def\etbl{\end{tabular}}
\def\btbb{\begin{tabbing}} \def\etbb{\end{tabbing}}
\def\beqar{\begin{eqnarray}} \def\eeqar{\end{eqnarray}}
\def\\{\hfill\break} \def\dit{\item{-}} \def\i{\item}
\def\bbb{\begin{thebibliography}{9} \itemsep=-1mm}
\def\ebb{\end{thebibliography}} \def\bb{\bibitem}
\def\bpic{\begin{picture}(260,240)} \def\epic{\end{picture}}
\def\akgt{\noindent{\bf Acknowledgements}}
\def\fgn{\noindent{\bf\large\bf Figure captions}}
\def\eps{\varepsilon}

\begin{document}
\begin{center}

{\Large
Two Specific Correlation Patterns as Indicators

for Various Random Multiplicative Cascade Processes }

\vskip 1cm

Wu Yuanfang\ \ \ Wang Yingdan\ \ \  Bai Yuting\ \ \ Liao Hongbo\ \ \ Liu Lianshou

\vskip 0.5cm

{\small \  Institute of Particle Physics, Huazhong Normal University, Wuhan 430079 China}

{\small FAX: 0086 27 87662646
\qquad email: wuyf@iopp.ccnu.edu.cn}
\date{ }
\end{center}

\begin{center}
\begin{minipage}{125mm}

\vskip 2cm
\begin{center}{\Large Abstract}\end{center}

\ \ \ \ It is suggested and demonstrated that
two specific 2-dimensional correlation patterns, fixed-to-arbitrary bin and
neighboring bin correlation patterns,
are efficient for identifying various random multiplicative cascade
processes. A possible application of these two correlation patterns to single event analysis in
Relativistic Heavy Ion Collider experiments is discussed.

\end{minipage}

\vs 1.5cm


\end{center}
\npg

It is well-known that nonlinear phenomena are very popular in 
nature. Once a set of anomalous scaling law of probability- or
``mass''-moments with respect to spatial size~\cite{multifractal}
is observed in experiment, we recognize that there is self-similar
multifractal structure. However, it is difficult to judge what kind
of mechanism most likely causes the results since the 
conventional measures have not provided sufficient information on 
underlying dynamics. This hinders us from further understanding and 
studying nonlinear physics. A useful measure for this purpose should 
not contain too much detail, but enough to capture the essence of
production mechanism. In this letter, as a first try along this
direction, we will suggest two specific correlation patterns which
manifest distinguishable characters of various random multiplicative 
cascade processes. They will be helpful in exploring the origin of 
nonlinear and correlation related phenomena.

A random multiplicative cascade process is the simplest example of
having a well-defined multifractal structure and has been used
extensively in various fields. For example, energy dissipation in
fully developed turbulence~\cite{multifractal,fluid} and successive
branchings in QCD parton-shower~\cite{QCD} are both this kind of
process and accordingly have multifractal structure. The main idea
of such a process is simply a series of self-similar random
cascades in spatial partitions. It is generally described as
follows: at the first step of the cascade, a given initial
interval with length $\Delta$ and unit probability $p_0=1$ is
split into $\tau$ subdivisions (``bins''). Usually, $\tau$ is an
integer with $\tau \geq 2$. The probability for a bin to be split
is determined by a weight $w$ which is distributed according to a
splitting function $p(w_1, w_2, \dots, w_\tau)$. Then each bin
obtained from the first step of the cascade is again independently
split into $\tau$ bins and so on. After $\nu=1, 2, ..., J$ steps,
the final number of bins is $\tau^{\nu}$, and the probability in a
specific bin $j_\nu$ is $\nu$ products of $w$ over all previous
generations: $p_{j_\nu}^\nu=w_{1j_1}w_{2j_2}\dots w_{\nu j_\nu}$,
where $j_\mu=1, 2, ..., \tau^\mu\ (\mu=1, 2, ..., \nu)$. The
distribution of $p_{j_\nu}^\nu$ obtained in this way fluctuates
violently from bin to bin, {\it cf.} Fig. 6
in~\cite{multifractal}.

By varying the splitting number and/or changing the splitting
function, all kinds of random multiplicative cascade processes can
be constructed. Especially, an additional physical constraint on
the splitting function, like probability conservation at each step
of splitting, will cause stronger correlations among the $\tau$
split bins than those without it. The smaller the splitting number
$\tau$ is, the stronger are the correlations among the ${\tau}$
bins. It is therefore necessary to investigate $n$-point
correlations in order to probe the differences among the various
random cascade processes. Let us start from the simplest 2-point
ones.

Why are conventional measurements for 2-point-correlation
moments~\cite{scaling} insufficient? This is because they are
usually defined at fixed correlation length with both horizontal
average over all bins in an event and longitudinal average over
all events in a sample. The fixed correlation length makes a
comparison between correlations with different scales impossible,
and the horizontal average smooths out the differences between
different pairs of bins. For these measurements, various random
multiplicative processes have the same kind of power law behavior
with diminishing correlation length, similar to the scale
invariance of ``mass'' moments. They have little help in the
identification of production mechanism. Hence, 2-point
correlation moments without horizontal averaging, but with various
correlation lengths, are necessary to be presented simultaneously.
It should be a {\it pattern} constructed by all kinds of 2-point
correlation moments.

A straightforward pattern of such a kind is the three dimensional
pattern of $2$-point correlation cumulants~\cite{Martin}, \beq
C_{K_1,K_2}=\langle \ln p_{K_1}\ln p_{K_2} \rangle_c = \langle \ln
p_{K_1}\ln p_{K_2} \rangle - \langle \ln p_{K_1} \rangle \langle
\ln p_{K_2} \rangle, \eeq \noindent where $K_1$ and $K_2$ are the
coordinates of two points. However, in this pattern the
differences among various random multiplicative cascade processes
are hidden in complicated background and hard to be observed, {\it
cf.} Fig.1, where the patterns for three different models, the
$\alpha$ model without probability conservation~\cite{scaling}, the
$p$ model with probability conservation~\cite{p} and the $c$ model
with both probability conservation and random probability
partition at each step of the splitting~\cite{c}, are presented.
The pattern manifests the common characteristics of random
multiplicative cascade processes, such as symmetry at each step of
splitting, the largest correlations of any bin with itself and so
on, but is insensitive to the basic model assumptions. Thus, it is
better to present the correlations of some specific two bins in a
two dimensional pattern so that the differences in various
processes are manifested.

A simple way to do so is to fix one bin $K_1$ and vary the left
one $K_2$. It is a fixed-to-arbitrary bin correlation pattern. It
contains correlation information at various lengths from the fixed
bin to all other bins, in which $\tau-1$ bins are separated from
the fixed one at the last cascade step and $\tau^J-\tau$ bins are
separated successively from the fixed one before the last step of
cascading. Another simple way is to fix the distance between the two
bins and change the position of them. Since the correlations among
$\tau$ bins at each step of splitting are important information,
the distance between two bins are  better to be as close to each
other as possible, {\it i.e.} $K_1=K$ and $K_2=K+1$, where $K=1,
2, ..., \tau^J-1$. This is a neighboring bin correlation pattern.
These two correlation patterns provide useful information on
cascade mechanism and filter out the common characteristics of
random multiplicative cascade processes.

In the following, as a support of above arguments, the
correlation cumulants for three different random cascade
processes, the $\alpha$, the $p$ and the $c$ models, are first
derived in order to show how the different appearances of these
cumulants are caused by the model assumptions. Then it is argued
and demonstrated why the measurements for these two patterns
provide information on the model assumptions. Finally, the
extension and application of the patterns are discussed.

These three models are examples of the random binary cascade
processes with symmetric splitting functions
$p(w_1,w_2)=p(w_2,w_1)$. In this case, Eq.(1) can be written
simply as~\cite{Martin} \beq
C_{K_1,K_2}=A(J-d_2)+B(1-\delta_{d_2,0}), \eeq where
$d_2=J-\sum_{j=1}^J\delta_{k_1^1,k_1^2}\cdots\delta_{k_j^1,k_j^2}$
is the ultrametric distance, which is a measure of how many
generations one has to move up before a common ancestor is found;
\beq
A={\partial^2Q[\lambda_1,\lambda_2]}/{\partial\lambda_1^2}|_{\lambda=0},
\qquad
B={\partial^2Q[\lambda_1,\lambda_2]}/{\partial\lambda_1\partial\lambda_2}|_{\lambda=0}
\eeq are respectively the so called ``same-lineage'' cumulant,
which is the correlation of the common ancestor of $K_1$ and
$K_2$, and the ``splitting'' cumulant, which measures the
correlation between the two parts split first from their common
ancestor; \beq Q[\lambda_1,\lambda_2]=\ln \left[ \int
dw_{1}dw_{2}p(w_1,w_2)\exp(\lambda_{1}
\ln{w_{1}}+\lambda_{2}\ln{w_{2}})\right] \eeq is the branching
generating function (BGF).

For the $\alpha$ model, the $w_1$ or $w_2$ at each step of
splitting is two possible numbers, $1+\alpha$ and $1-\alpha$, with
equal probability for each of them. Hence, its splitting
function~\cite{Martin} can be written as: \beq p(w_1,w_2)=\frac
{1}{4}\biggl[\delta\biggl(w_1-(1+\alpha)\biggr)+\delta\biggl(w_1-(1-\alpha)\biggr)\biggr]
\biggl[\delta\biggl(w_2-(1+\alpha)\biggr)+\delta\biggl(w_2-(1-\alpha)\biggr)\biggr],
\eeq \noindent where $\alpha$ is a fixed parameter in the region
[-1, 1]. Inserting this function into Eq.(4) and then Eq.(4)
into Eq.(3),  the ``same-lineage'' and the ``splitting'' cumulants
can be easily derived: \beq A_\alpha=\frac{1}{4}\biggl[\ln
\frac{1+\alpha}{1-\alpha}\biggr]^2, \ \ \  B_\alpha=0. \eeq
\noindent $B_\alpha=0$ means that there is no correlation between
the two parts for every splitting. This is due to the independence
of $w_1$ and $w_2$ assumed in the model.

For the $p$ model, the summation of $w_1$ and $w_2$ keeps to be
$2$, with $w_1=1\pm\beta$ and $w_2= 1\mp\beta$ as the consequence
of probability conservation, here $\beta$ is a fixed parameter in
the region [-1, 1]. Its splitting function~\cite{Martin} is \beq
p(w_1,w_2)=\frac
{1}{2}\biggl[\delta\biggl(w_1-(1+\beta)\biggr)+\delta\biggl(w_1-(1-\beta)\biggr)\biggr]
\delta(w_1+w_2-2), \eeq \noindent and its ``same-lineage'' and
``splitting'' cumulants are \beq A_p=\frac{1}{4}\biggl[\ln
\frac{1+\beta}{1-\beta}\biggr]^2, \ \ \ B_p=-\frac{1}{4}\biggl[\ln
\frac{1+\beta}{1-\beta}\biggr]^2. \eeq \noindent Due to
probability conservation, the ``splitting'' cumulant $B$ has sign
opposite to the ``same-lineage'' cumulant $A$. Moreover, since
$\beta$ is a unique fixed number, the derivatives of BGF with
respect to one of the two $\lambda$'s twice are equivalent to
those with respect to both of them. This is why the absolute value
of the ``same-lineage'' cumulant $A$ is equal to that of the
``splitting'' cumulant $B$.

For the third $c$ model,  $w_1+w_2=1$ similar to the $p$ model.
Distinct from the $p$ and the $\alpha$ models, the weight $w$ is
allowed to be a random number {\it e.g.} $ w_1={(1+\gamma r)}/{2}$
and $w_2={(1-\gamma r)}/{2}$, here $r$ is a random number
uniformly distributed in the interval $[-1,1]$ and $\gamma$ is a
fixed model parameter in the region $[0,1]$.  Obviously, this
model is more flexible and closer to the real system with
probability conservation. The splitting function is: \beq
p(w_1,w_2)=\frac{1}{\gamma}\biggl[\theta\biggl(w_1-\frac{1-\gamma}{2}\biggr)-
\theta\biggl(w_1-\frac{1+\gamma}{2}\biggr)\biggr]\delta
(w_1+w_2-1). \eeq and its ``same-lineage'' and ``splitting''
cumulants are: \beq A_c=\frac{1}{\gamma}
\int_{\frac{1-\gamma}{2}}^{\frac{1+\gamma}{2}} (\ln
w_1)^2dw_1-\frac{1}{\gamma^2}
\biggl[\int_{\frac{1-\gamma}{2}}^{\frac{1+\gamma}{2}}\ln
w_1dw_1\biggr]^2, \nonumber \eeq \beq B_c=\frac{1}{\gamma}
\int_{\frac{1-\gamma}{2}}^{\frac{1+\gamma}{2}} \ln w_1 \ln
(1-w_1)dw_1-\frac{1}{\gamma^2}\biggl[\int_{\frac{1-\gamma}{2}}^{\frac{1+\gamma}{2}}\ln
w_1dw_1\biggr]^2. \eeq The ``splitting'' cumulant is nonzero due to
probability conservation. The absolute values of the
``same-lineage'' and the ``splitting'' cumulants are unequal since
$w_1$ and $w_2$ are no longer fixed to a particular number, unlike
the $p$ model.

From the above analytic derivation, we see that the three random
cascade models are very different in the absolute values of their
``splitting'' cumulants and in the relative values of these
cumulants to the ``same lineage'' cumulants. These differences 
are due to the basic model assumption and 
independent of the model parameters.
Therefore, a useful 2-point correlation pattern should contain
information on the ``splitting'' cumulant and its relation to the
``same lineage'' cumulant. This requirement turns out to be two
possible combinations of the coefficients $A$ and $B$ in Eq.(2):
(1) $C_{K_1,K_2}=B$, {\it i.e.} $J-d_2=0$ and
$1-\delta_{d_2,0}=1$; (2) $C_{K_1,K_2}=A+B$, {\it i.e.} $J-d_2=1$
and $1-\delta_{d_2,0}=1$; for the first combination, $d_2=J$ and
$d_2\not=0$ means that common ancestor has to be at the initial
interval if $J > 0$. The fixed-to-arbitrary bin correlation
pattern happens to have $2^{J-1} \cdot (2-1)$ pairs of this kind
of bins. The neighboring bin correlation pattern also has a pair
of this kind of bins at center $K=2^{J-1}$. For the second
combination, $d_2=J-1$ and $d_2\not=0$ implies that common
ancestor locates at the first cascade step. In the
fixed-to-arbitrary bin correlation pattern, $2^{J-2} \cdot (2-1)$
pairs of bins have common ancestor at the first cascade step. The
neighboring bin correlation pattern again has two pairs of such a
kind of bins at $K=2^{J-2}$ and  $K=2^{J-2}\cdot 3$ respectively.
Hence the suggested two correlation patterns are useful measures
in identifing the above mentioned three random cascade processes.

We first plot the fixed-to-arbitrary bin correlation pattern for
the $\alpha$, the $p$ and the $c$ models in Figs. 2(a), (b) and
(c), respectively, with the model parameters $\alpha=0.4$,
$\beta=0.4$, $\gamma=0.8$, $J=6$. The ordinate is the correlation
strength $C_{1,K_2}$; the abscissa is the bin $K_2$ which varies
as $K_2=2, 3, ..., 64$, and the fixed bin is located at $K_1=1$.
Since the larger the distance between the two bins is, the earlier
the two bins are separated in cascade process, the correlation
strength decreases as the arbitrary bin $K_2$ goes further away
from the fixed one $K_1$. The number of points with the same
height or same correlation strength increases as  $2^{\nu-1} \cdot
(2-1)$ with $\nu=1, 2, ..., J$. The number $2$ comes from the
binary cascade. Generally, it is $\tau^{\nu-1}\cdot (\tau-1)$.
So by counting the number of points in each
correlation strength, the splitting number of the cascade process
can be estimated.

The possible correlation strengths in these three models are
different: (1) the weakest one in the $\alpha$ model is zero; (2)
there are both zero and negative ones in the $p$ model; (3)
negative ones, but no zero ones, are allowed in the $c$ model.
These differences come from the different model assumptions. The
absence of negative correlations implies independent splitting in
the model,  $B=0$, such as is the case for the $\alpha$ model. If
there are both negative and zero correlations in the pattern, the
positive one from the ``same-lineage'' cumulant $A$ must be equal
to the negative one from the ``splitting'' cumulant $B$. So that
$B < 0$ for the weakest one and $A+B=0$ for the following one, as
for the $p$ model.  Having negative, but no zero, correlation,
means that the ``same-lineage'' cumulant is different from the
``splitting'' cumulant, or $A+B\not= 0$ and $B < 0$, as for the
$c$ model.

The neighboring bin correlation pattern $C_{K,K+1}$ with $K=1, 2,
..., 2^J-1$ can show us the same thing. In Fig.3, the patterns for
the three models are presented. Here, the strongest correlations
are of those two neighboring bins split at the last step of
cascade, {\it i.e.}, $K=1, 3,..., 2^J-2^0$; the next ones happen
to be of those split before the last step of cascade, {\it i.e.},
$K=2, 6, 10, ..., 2^J-2^1$; and so on. The number of points at a
certain correlation strength is determined by $2^{\nu-1}$, or
$\tau^{\nu-1}$ in general, with $\nu=J, J-1, ..., 1$. There are
only positive and zero correlation spectra in the $\alpha$ model.
In the $p$ model, the correlation spectrum ranges over positive,
zero and negative values, while for the $c$ model, only positive
and negative correlations are allowed with no zero ones.

Therefore, from the measurements of these two correlation
patterns, we can find out by what kind of random multiplicative
cascade process the observed system are produced. This is a
first step toward the revelation of underlying mechanism of
multifractal phenomena. The extension of these two patterns to the
mixture of different random cascade processes and other mechanisms,
which can lead to multifractal structure, is obviously the next
step. Their corresponding patterns will not be so regular as those
of the above mentioned three processes. For the further
identification of the production mechanism in general, other
measures may be needed in addition to these two. Nevertheless,
having established these two patterns for some known mechanisms,
it is interesting to compare the patterns measured in practice
with them and judge if there are similar generating mechanism.

A possible application of these two correlation patterns is in the
single event analysis of Relativistic Heavy Ion Collider (RHIC)
experiments~\cite{rhic}. In these experiments, the probability of
final state particles falling into a particular bin in rapidity or
pseudo-rapidity can be well estimated by~\cite{wsfs}, $ p_{K_i}
\approx {n_{K_i}}/{N}$, where $n_{K_i}$ is the number of final
state particles falling into the $K_i$-th bin and $N$ is the total
number of particles in the event, which is typically several
thousand~\cite{rhicn}. At the current experimental resolution in
rapidity, there is almost no empty bin. In this case, it is much
better to investigate directly the pattern of a single event
without averaging over all events in the sample. The 2-point
correlation cumulant for a single event can be defined as: \beq
C_{K_1,K_2}^{(\rm e)}=\ln p_{K_1}\ln p_{K_2}-\langle \ln
p_{K_1}\rangle \langle \ln p_{K_2}\rangle, \eeq its average over
all events in a sample equals to the 2-point correlation cumulant
$C_{K_1,K_2}$ of Eq.(1). The whole event sample can then be
classified by the measured correlation patterns of single event.
Some events which have undergone a phase transition to Quark Gluon
Plasma (QGP) will most likely have very special patterns, since
during this transition, all kinds of correlations, the long as
well as the short ones, will change dramatically.

In this letter, we investigate two specific 2-dimensional
correlation patterns: the fixed-to-arbitrary bin and the
neighboring bin correlation patterns. The former shows
correlations at various lengths, and the latter has short-range
correlations at different positions. They are sensitive to the
underlying mechanisms and work equally well in
identifying various random multiplicative cascade processes. As an
example, their application to the analysis of single event of
current Relativistic Heavy Ion Collider(RHIC) experiments has been
discussed. For more-than-2-point correlations, we can also find
some useful correlation patterns in a similar way. The
measurements for all of them will hopefully open a new world in
the research of nonlinear and correlation related physics.

\vs 3mm

We are grateful for the helps of Dr. Jimmy MacNaughton in the
presentation of the paper and the good suggestions of Dr. Nu Xu.
This work is supported in part by the SFC under project 90103019;
Cross-century Talent Foundation of National Education Committee of
China under Grant No.1 [1998].

\newpage
\def\Journal#1#2#3#4{{#1} {\bf #2} #3 (#4)}
\def\CSB{\em Chinese Science Bulletin}
\def\NCA{\em Nuovo Cimento} \def\NIM{\em Nucl. Instrum. Methods}
\def\NIMA{{\em Nucl. Instrum. Methods} A} \def\NPB{{\em Nucl. Phys.} B}
\def\PLB{{\em Phys. Lett.}  B} \def\PRL{\em Phys. Rev. Lett.}
\def\PRD{{\em Phys. Rev.} D} \def\ZPC{{\em Z. Phys.} C}
\def\PRE{{\em Phys. Rev.} E} \def\PRC{{\em Phys. Rev.} C}
\vs 1cm

\newpage
\ni{\Large\bf Figure Captions}

\vs 0.5cm {\pa=0pt{\ls=15mm\rightskip0mm \hs-12mm {\bf Fig.1} \
The correlation cumulants $C_{K_1,K_2}$ of
 ($a$) the $\alpha$ model,
($b$) the $p$ model and the ($c$) $c$ model.
\par}}

\vs 0.5cm {\pa=0pt{\ls=15mm\rightskip0mm \hs-12mm {\bf Fig.2} \
The fixed-to-arbitrary bin correlation patterns $C_{1,K_2}$ of
($a$) the $\alpha$ model, ($b$) the $p$ model and ($c$) the $c$
model, where the ordinate is correlation strength $C_{1,K_2}$ and
the abscissa is $K_2=2, 3, \cdots, 64$. The solid lines in the
figures indicate zero correlation strength.
\par}}

\vs 0.5cm {\pa=0pt{\ls=15mm\rightskip0mm \hs-12mm {\bf Fig.3} \
The neighboring bin correlation patterns $C_{K, K+1}$ of ($a$) the
$\alpha$ model, ($b$) the $p$ model and ($c$) the $c$ model, where
the ordinate is correlation strength  $C_{K, K+1}$ and the
abscissa is $K=1,2,\cdots, 63$. The solid lines in the figures
indicate zero correlation strength.
\par}}

\newpage
\begin{center}
\begin{picture}(300,300)
\put(-140,-125)
{
{\epsfig{file=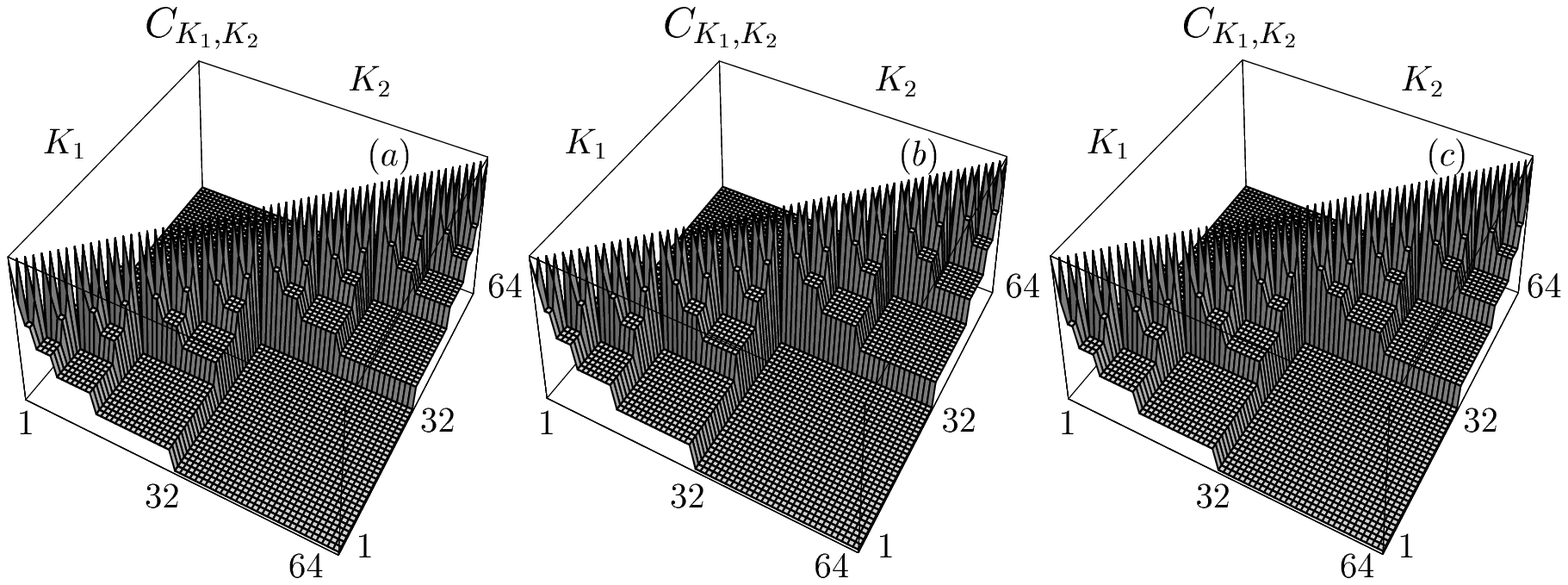,width=500pt,height=600pt}}
}

\put(-140,-580)
{
{\epsfig{file=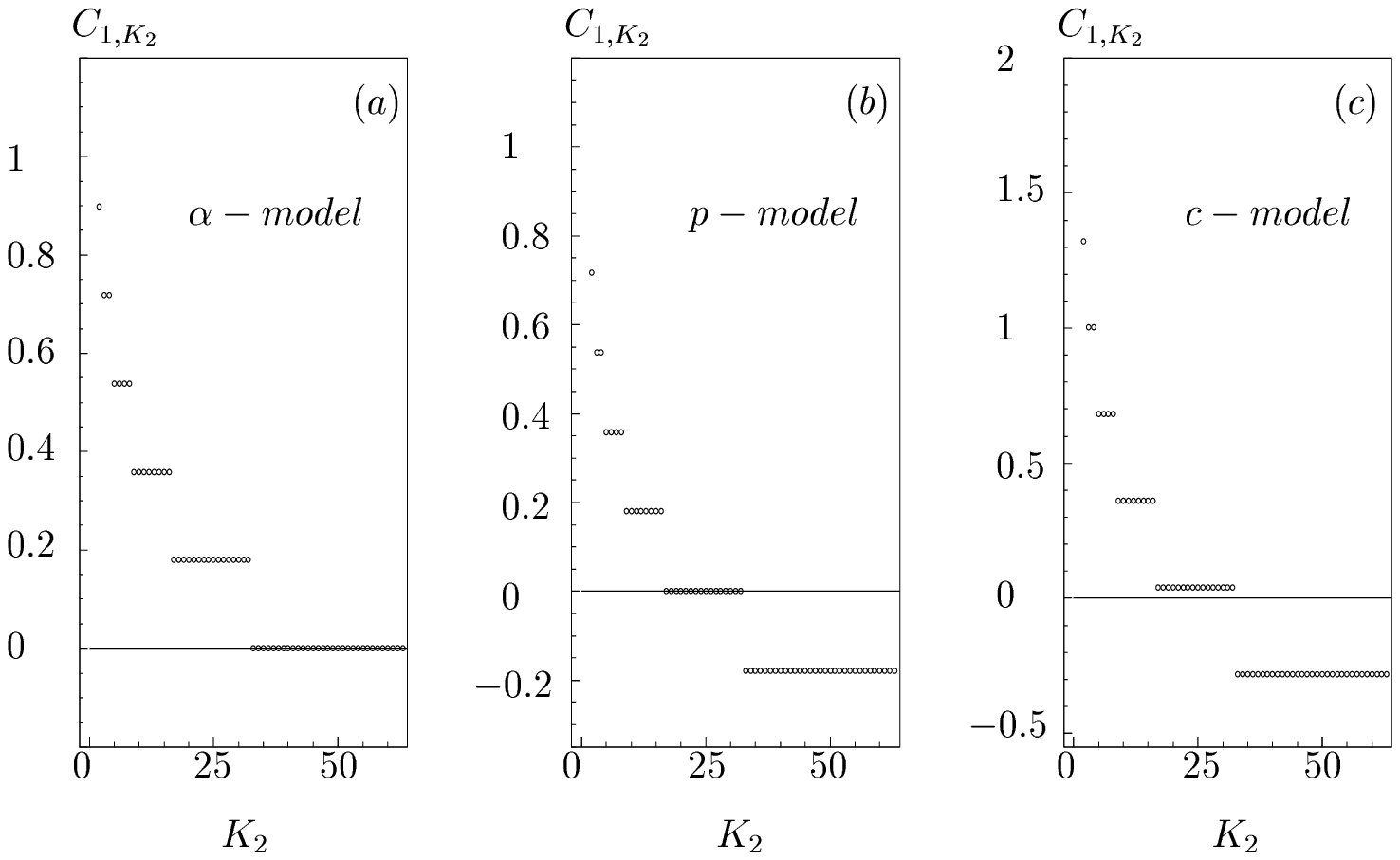,width=550pt,height=700pt}}
}
\end{picture}
\end{center}

\vs -4.0cm
\n{\hskip 9.5cm {\large Fig.1}}
\vs 10.cm
\n{\hskip 9.5cm {\large Fig.2}}

\newpage
\begin{center}
\begin{picture}(300,300)
\put(-100,-250)
{
{\epsfig{file=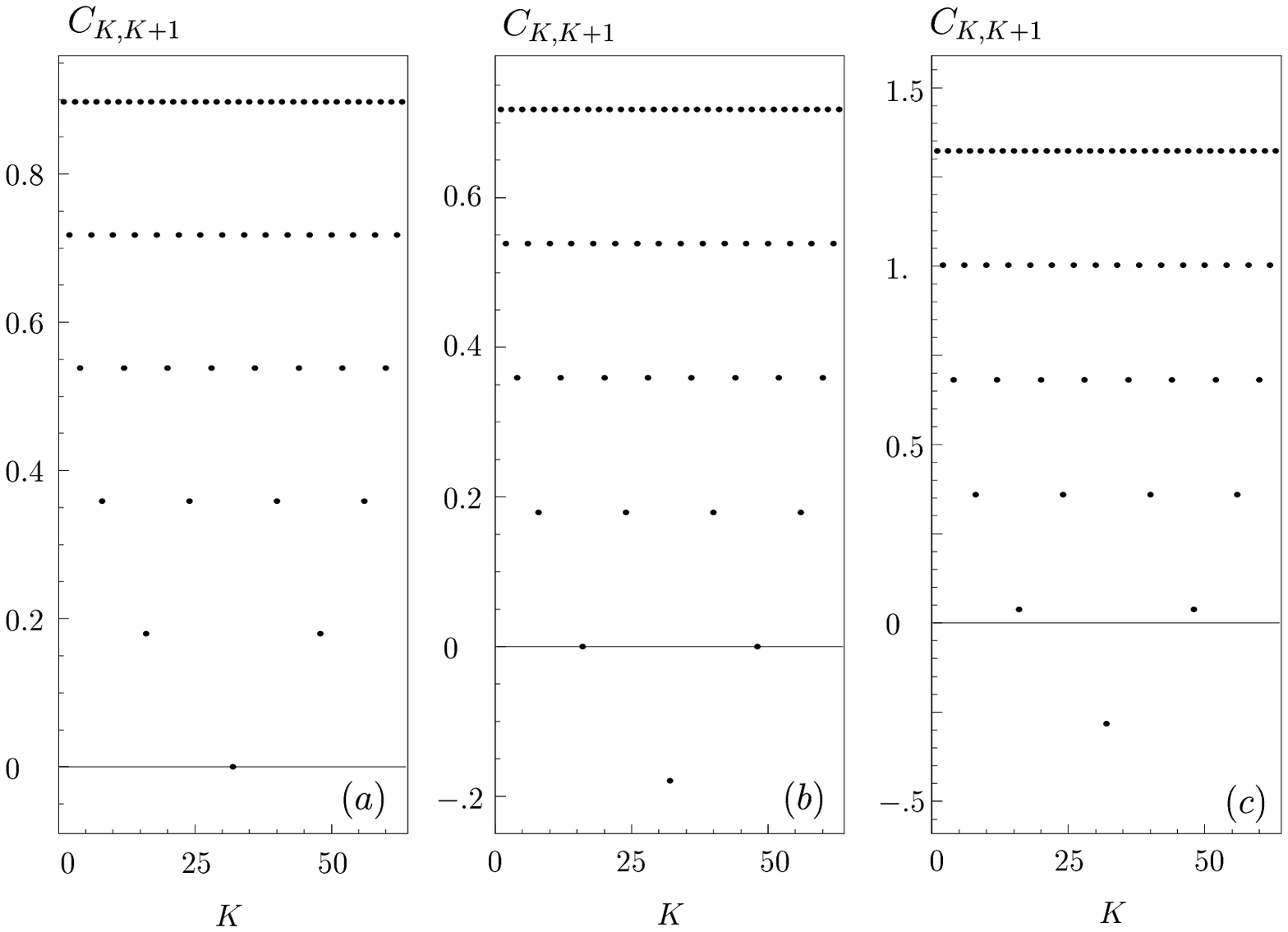,width=500pt,height=600pt}}
}
\end{picture}
\end{center}

\vs 5.0cm
\n{\hskip 10cm {\large Fig.3}

\end{document}